\documentclass[aps,prl,twocolumn,showpacs]{revtex4}
\usepackage{epsfig}
\usepackage{color}
\begin{document}

\title{Mott insulators and correlated superfluids in ultracold
  Bose-Fermi mixtures} 
\author{F. H\'ebert$^1$, F. Haudin$^1$, 
L.~Pollet,$^2$
and 
G.G. Batrouni$^1$}
\affiliation{$^1$INLN, Universit\'e de Nice-Sophia Antipolis, CNRS; 
1361 route des Lucioles, 06560 Valbonne, France}
\affiliation{$^2$Institut f\"ur Theoretische Physik, ETH Z\"urich,
  CH-8093, Switzerland}

\begin{abstract}
We study the effects of interaction between bosons and fermions in a
Bose-Fermi mixtures loaded in an optical lattice.  We concentrate on
the destruction of a bosonic Mott phase driven by repulsive
interaction between bosons and fermions.  Once the Mott phase is
destroyed, the system enters a superfluid phase where the movements of
bosons and fermions are correlated.  We show that this phase has
simultaneously correlations reminiscent of a conventional superfluid
and of a pseudo-spin density wave order.
\end{abstract}

\pacs{
05.30.Jp, 
03.75.Hh, 
71.10.Fd, 
02.70.Uu  
}
\maketitle

Cold atoms loaded on optical lattices provide a new tool to study
condensed matter systems.  These systems are experimental realizations
of lattice models for quantum interacting particles such as Hubbard
models. They provide a simple and controllable way to study various
interesting phenomena such as the superfluid to Mott insulator (MI)
transition \cite{Greiner} or Anderson localization \cite{local}.
Recently, there has been growing interest in the physics of
boson-fermion mixtures at low temperature.  These mixtures served
originally to lower the temperature of the fermions by sympathetic
cooling \cite{sympacooling} with the bosons in order to study the low
temperature physics of interacting fermions, such as superfluidity in
the BCS or BEC regime \cite{Ketterle2}.

In some cases the interactions between fermions and bosons cannot be
neglected.  Fermions have, for example, been shown to reduce the phase
coherence of weakly interacting superfluid bosons \cite{bosefermiexp}.
Theoretical studies \cite{Svistunov,Lewenstein1, Demler} have
suggested the possibility for the system to form pairs (polarons) of
bosons and fermions (particle-particle or particle-hole), with these
composite fermionic pairs then adopting different phases such as
normal Fermi liquid or solid phases.  This system was studied
numerically in one dimension by Sengupta et Pryadko \cite{Pinaki1} and
shows a commensurate filling phase where the sum of the number of
bosons, $N_{\rm b}$, and fermions, $N_{\rm f}$, is equal to the number
of lattice sites $L$: $N_{\rm b} + N_{\rm f} = L$.  This phase has
been studied simultaneously in the special case where $N_{\rm
b}=N_{\rm f}=L/2$ by Pollet {\it et al.}  \cite{Pollet1} who showed
that it could be understood, in the strong interaction limit, as the
pseudo-spin density wave (PSDW) phase previously proposed by Kuklov
and Svistunov \cite{Svistunov}.  This kind of system has also been
studied in the presence of harmonic traps \cite{trap}, but the
comparison with experiment~\cite{bosefermiexp} is problematic.

In this letter we study the effect of added fermions on a bosonic MI
phase \cite{Greiner} at low temperature. As the boson-fermion
interaction is increased, the MI is destroyed and the system enters a
superfluid phase (SF). We will review the properties of this phase and
show that it can be crudely described as composed of superfluid bosons
and of the aboved mentionned PSDW phase. In addition, we study the
properties of the PSDW phase when there is an imbalance in the boson
and fermion populations but with $N_{\rm b} + N_{\rm f} = L$.

\begin{figure}[!t]
\epsfig{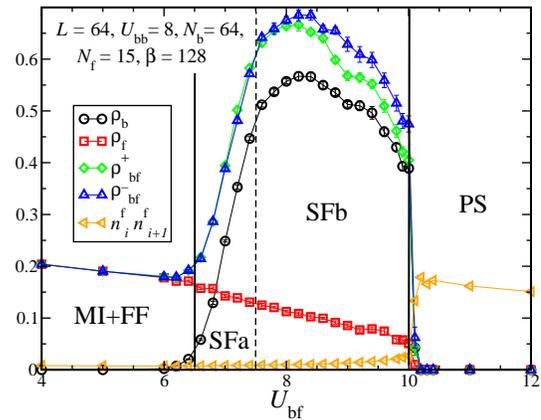}
\caption{At low $U_{\rm bf}$, the system consists of free fermions on
a bosonic Mott insulator substrate. As $U_{\rm bf}$ becomes larger the
Mott insulator is destroyed and the system becomes superfluid
(SF). There is a crossover between two different superfluid regions
(see the text). Finally for large values of $U_{\rm bf}$, fermions and
bosons phase separate (PS).}
\label{destroymott}
\end{figure}

The system is described by a one-dimensional repulsive Hubbard model
\begin{eqnarray}
  \label{Hamiltonian}
\mathcal H&=& -t\sum_{i} \left(b_{i}^\dagger b_{i+1} + f_{i}^\dagger
       f_{i+1} + {\rm h.c.} \right) \label{hamiltonian} \\ &+& U_{\rm
       bb} \sum_{i} {n}^{\rm b}_{i} (n^{\rm b}_{i}-1) / 2 + U_{\rm bf}
       \sum_{i} n^{\rm b}_{i} n^{\rm f}_{i}, \nonumber
\end{eqnarray}
where the operator $b^\dagger_i$ ($b_i$) creates (destroys) a boson on
site $i$ while $f^\dagger_i$ and $f_i$ are the corresponding fermionic
operators. The first term in (\ref{hamiltonian}) describes the hopping
of bosons and fermions between neighboring sites. We have chosen the
hopping amplitude $t=1$, which sets the energy scale, to be the same
for both species.  $n^{\rm b}_{i}$ ($n^{\rm f}_{i}$) is the bosonic
(fermionic) number operator at site $i$ and $U_{\rm bb}$ and $U_{\rm
bf}$ are the boson-boson and boson-fermion contact interaction terms.

The bosonic, $G_{\rm b}(x) = \langle b^\dagger_i b_{i+x} \rangle$,
fermionic, $G_{\rm f}(x) = \langle f^\dagger_i f_{i+x} \rangle$, and
composite, $G_{\rm bf}(x) = \langle b^\dagger_i f^\dagger_{i+x}
b_{i+x}f_{i} \rangle$, Green functions, give access to the phase
correlations of bosons, fermions and particle-hole pairs,
respectively.  A non vanishing bosonic superfluid density $\rho_{\rm
b} = {\langle W^2_{\rm b}\rangle/2 \beta }$ or a fermionic stiffness
$\rho_{\rm f} = \langle W^2_{\rm f}\rangle / 2 \beta $ shows the
presence of slowly (algebraically) decaying phase correlations
($W_{\rm b}$ and $W_{\rm f}$ being the bosonic and fermionic winding
numbers).  The paired (+) and counter-rotating (--) superfluid
densities, $\rho^\pm_{\rm bf} = \langle \left(W_{\rm b}\pm W_{\rm
f}\right)^2\rangle/2 \beta$ are both equal to $\rho_{\rm b} +
\rho_{\rm f}$ if the movements of the two types of particles are
uncorrelated and differ if the movements are correlated (or
anticorrelated).

We study this model using a canonical quantum Monte Carlo method based
on ``worm" updates which is extremely efficient for the measurements
of equal time Green functions \cite{algo1}.  This method was modified
to tackle the case of mixtures by allowing simultaneous moves of the 
bosons and fermions \cite{algo3}.

\begin{figure}[!t]
\epsfig{figure=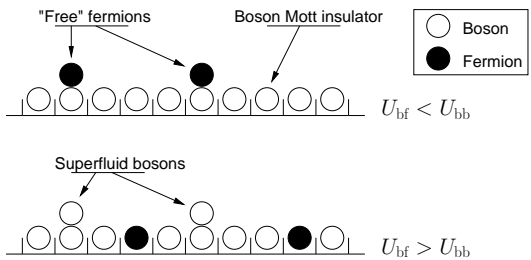,width=7cm}
\caption{Schematic representation of the destruction of the bosonic
Mott insulator. As $U_{\rm bf}$ is increased, the sites that were
occupied by a boson and a fermion are replaced by sites occupied by
two bosons. The number of bosons per site is no longer 1 and the
system can become superfluid.}
\label{shema}
\end{figure}

We set the number of bosons $N_{\rm b}$ to be equal to the number of
sites $L=64$ and set $U_{\rm bb}=8$, to obtain a solid Mott phase for
the bosons in absence of interactions with the fermions.  We studied
the evolution of the system as $U_{\rm bf}$ is gradually increased for
two different values of the number of fermions ($N_{\rm f} = 15$ and
35).

\begin{figure}[!t]
\epsfig{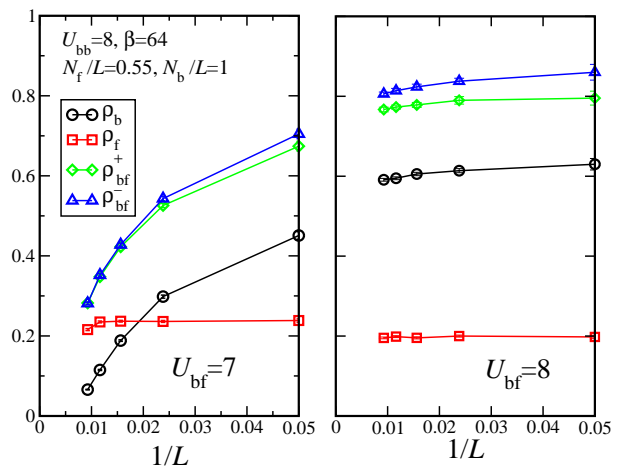}
\caption{Finite size scaling of different densities for two values of
the interaction. The SFa regime observed in Fig.~\ref{destroymott}
around $U_{\rm bf} \simeq 7$ disappears for larger size whereas the
SFb regime persists for $U_{\rm bf} \simeq 8 $.}
\label{ffs}
\end{figure}

For $U_{\rm bf} < 6.5$ the superfluid densities
(Fig. \ref{destroymott}) show that the system behaves as in the non
interacting limit (MI+FF).  The bosons form a MI where $\rho_{\rm
b}=0$, each site is occupied by only one boson and jumps are forbidden
because they cost a large energy $U_{\rm bb}$. On the contrary, the
fermions move freely as they experience the same interaction energy
$U_{\rm bf}$ on all the sites ($\rho_{\rm f} \ne 0$).  When $U_{\rm
bf}$ becomes larger than $U_{\rm bb}$ the Mott insulator is destroyed
and the bosons become superfluid. In this case, it is energetically
favorable to form pairs of bosons on the same site and to have
isolated fermions (Fig. \ref{shema}).  The system is then split into
doped superfluid bosonic regions and fermions.  If the repulsion
between bosons and fermions is not too strong ($6.5 < U_{\rm bf} <
10$), the bosons can pass from one superfluid region to the next one
by the tunnelling through fermion-occupied sites and a superfluid (SF)
phase is established. For $U_{\rm bf} > 10$ the tunnelling is
suppressed and there is a phase separation between independent
superfluid regions separated by fermionic regions.  This case is
difficult to study numerically because the system is frozen.

When the MI is destroyed, two distinct regions \cite{LL}, separated by
a crossover, appear: a first region (SFa, $6.5 < U_{\rm bf} < 7.5$)
where correlations of the movements of bosons and fermions are not
visible ($\rho_{\rm b} + \rho_{\rm f} = \rho^\pm_{\rm bf}$) and a
second one (SFb, $7.5 < U_{\rm bf} < 10$) where $\rho^-_{\rm bf} >
\rho^+_{\rm bf}$.  In this latter region moving a fermion
independently of the surrounding bosons costs $U_{\rm bf}$, while
exchanging a boson and a fermion has no energy gap associated. These
favored exchanges lead to the observed anticorrelation of winding
numbers.

The SFa region disappears in the thermodynamic limit (or at least
becomes very narrow) as can be observed in Fig.~\ref{ffs}, where
$\rho_{\rm b}$ goes to zero with $1/L$ for $U_{\rm bf}=7$. On the
contrary, around $U_{\rm bf}=8$, the SFb phase persists in the large
size limit.  We then expect a direct transition from the MI+FF phase
to the SFb regime (Fig.~\ref{ffs} corresponds to a density of fermions
$N_{\rm f} / L \simeq 0.55$ but the two values of $U_{\rm bf}$
correspond to the same phases as in Fig.~\ref{destroymott} where
$N_{\rm f} / L \simeq 0.23$).

\begin{figure}[ht]
\epsfig{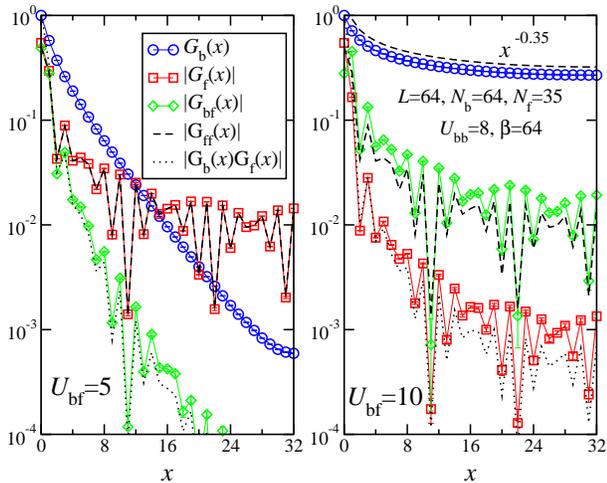}
\caption{Left : For $U_{\rm bf}=5$ the bosons are in a Mott insulating
state and $G_{\rm b}$ decays exponentially. The fermions follow the
free fermions behavior. Right: In SFb regime, we observe phase
coherence in all the cases. $G_{\rm bf}$ is larger than $G_{\rm f}$.}
\label{green}
\end{figure}

In Fig.~\ref{green}, we show the Green functions in the MM+FF (left)
and in the SFb (right) phases. In the Mott phase, $G_{\rm b}(x)$
decays exponentially while $G_{\rm f}(x)$ decays algebraically
following the free fermion Green function $G_{\rm ff}(x) = \sin(N_{\rm
f} \pi x/L)/\sin(\pi x /L)$.  The two species are independent, in
spite of the large value of $U_{\rm bf}$, and consequently $G_{\rm
bf}=G_{\rm b}G_{\rm f}$.

In the SFb, all the Green functions decay algebraically. The slowest
decay is obtained for the bosons with $G_{\rm b} \propto \sin(\pi
x/L)^{-0.35}$ while $G_{\rm f} \propto \sin(N_{\rm f}\pi x/L) \cdot
\sin (\pi x /L)^{-1.5}.$ In this case, $G_{\rm bf}$ is much larger
than both the product $G_{\rm b}G_{\rm f}$ and $G_{\rm f}$ and follows
(approximately) $G_{\rm ff}$.  This behavior confirms that
boson-fermion pairs are important degrees of freedom in this system
(in the SFa region, on the contrary, $G_{\rm f}$ is larger than
$G_{\rm bf}$ \cite{LL}).

\begin{figure}
\epsfig{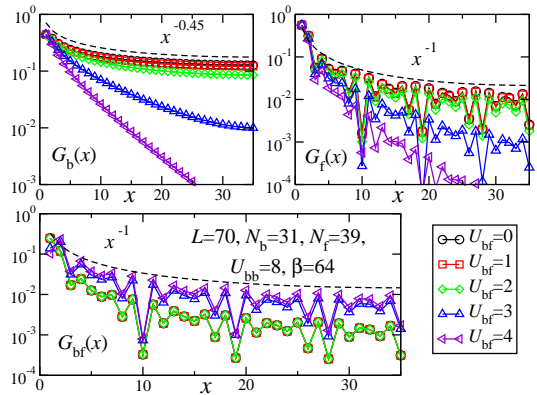}
\caption{Green functions for several $U_{\rm bf}$.  At low
interaction, there are superfluid bosons and mostly free fermions
(Luttinger liquids). At large interaction $U_{\rm bf} > 3$, the system
enters the PSDW regime, individual $G_{\rm b}$ and $G_{\rm f}$ decay
exponentially while $G_{\rm bf}$ remains quasi long ranged. }
\label{3G}
\end{figure}

\begin{figure}
\epsfig{figure=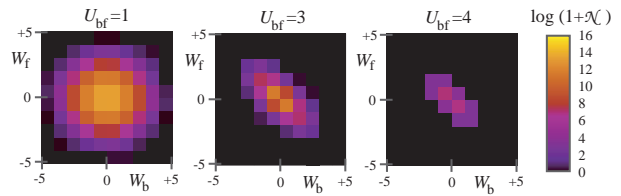,width=8cm}
\caption{Histograms of the winding numbers ${\cal N}(W_{\rm b},W_{\rm
f}$). Same parameters as in Fig.~\ref{3G}.  In a low interaction
Luttinger liquid phase ($U_{\rm bf}=1$) $W_{\rm b}$ and $W_{\rm f}$
are not correlated and ${\cal N}$ is isotropic (left panel).  With
increasing $U_{\rm bf}$ the winding numbers become correlated (middle
panel), while in the PSDW phase ($U_{\rm bf} = 4$), ${\cal N}$ is
strongly anisotropic and only allows the cases where $W_{\rm b} = -
W_{\rm f}$ (right panel).}
\label{histo}
\end{figure}

Returning to Fig.~\ref{shema}, an intuitive but crude physical
interpretation is, that our system can be roughly described as
composed of $N_{\rm f}$ superfluid bosons and a layer composed of
$N_{\rm f}$ fermions and $L-N_{\rm f}$ bosons.  For such a ``filled"
layer, the description of the system as pseudo-spins is possible
\cite{Pollet1, Svistunov} whose properties we now review.  Once again,
we concentrate on the case where the bosons are strongly interacting
($U_{\rm bb} = 8$) and all the cases where $N_{\rm f}+N_{\rm b} = L$.
First, for small $U_{\rm bf}$, the full system is composed of two
weakly interacting Luttinger liquids, one for each species. Green
functions decay algebraically (Fig.~\ref{3G}, $U_{\rm bf} \le 2$) and
the winding numbers $W_{\rm b}$ and $W_{\rm f}$ are not correlated
(Fig.~\ref{histo}, left).

For large enough $U_{\rm bf}$ the system enters a PSDW phase: The
interactions forbid double occupancy of a site by two bosons or by a
fermion-boson pair and a pseudo-spin along the $z$-axis on site $i$
may be defined as $\sigma_i^z= n^{\rm b}_i - n^{\rm f}_i = \pm 1$.
The model can then be mapped on the XXZ Hamiltonian
$$\mathcal{H}_{\rm XXZ} = J_{xy} \left( \sigma_i^x \sigma_{i+1}^x +
\sigma_i^y \sigma_{i+1}^y \right) + J_{z} \sigma_i^z \sigma_{i+1}^z$$
where $J_{xy} = - t^2 / U_{\rm bf}$ and $J_z = t^2 ( 1 / U_{\rm bf} -
1 / U_{\rm bb})$.  The magnetic correlations along the $xy$ axes
correspond to particle-hole correlations $G_{\rm bf}(x)$ as
$\sigma^+_i \propto b^\dagger_i f_i$.  Bosonic and fermionic Green
functions decay exponentially as the hopping of individual particles
is energetically suppressed (Fig.~\ref{3G}, $U_{\rm bf}\ge 3$).  The
hopping of bosons and fermions and, consequently, the winding numbers
are strongly anti-correlated $W_{\rm b} = - W_{\rm f}$
(Fig.~\ref{histo}, right) which leads to $\rho_{\rm b} = \rho_{\rm f}
= \rho^-_{\rm bf} / 4$ and $\rho^+_{\rm bf} = 0 $.

Along the $z$ axis, one defines a spin correlation function $C_{\rm
psdw}(x) = \left\langle \sigma^z_i \sigma^z_{i+x} \right\rangle
$. Depending on the value of $J_z$, the system favors
antiferromagnetic (AF), for $U_{\rm bf} < U_{\rm bb}$, or
ferromagnetic (FM), for $U_{\rm bf} > U_{\rm bb}$, correlations along
the $z$-axis.  For $U_{\rm bf} > 2 U_{\rm bb}$, $|J_z| > |J_{xy}|$,
the dominant correlations are ferromagnetic along the $z$-axis which
leads to phase separation for the bosons and fermions.  In all other
cases, the $xy$ term is dominant, {\it i.e.}, $G_{\rm bf}$ is the
correlation function that has the slowest decay.

\begin{figure}
\epsfig{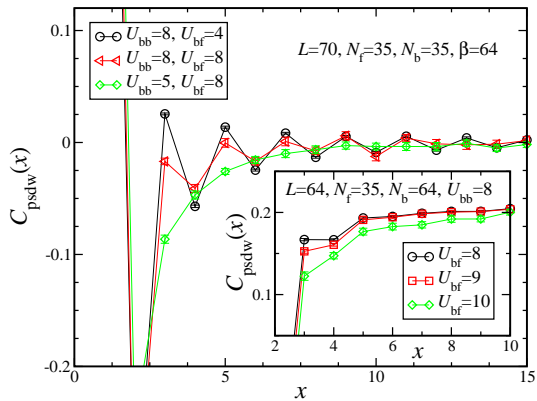}
\caption{Pseudo spin correlation function $C_{\rm psdw}(x)$ in the
case $N_{\rm f} = N_{\rm b} = L / 2 $ for the AF (circles), pure-$xy$
(triangles) and FM (diamonds) regimes. Inset: $C_{\rm psdw}(x)$ in the
CSF phase for different values of $U_{\rm bf}$. The AF signal
decreases as expected.}
\label{Sk}
\end{figure}

We concentrate on this latter case. In the AF regime along the $z$
direction, we observe the characteristic oscillations around zero, the
FM correlations decreasing much more rapidly than the AF ones (see
Fig. \ref{Sk}).  The presence of defects in the AF due to the
imbalance between the populations of up and down spins leads to
oscillations with beating where $C_{\rm psdw}(x)$ behaves like
$\cos(2\pi N_{\rm f}x/L)$ multiplied by a power law envelope.  Density
correlation functions exhibit similar beating oscillations,
reminiscent of the solitonic excitations observed in the extended
Hubbard model \cite{soliton}.  The commensurate oscillations in
$\cos(\pi x)$ are recovered only for $N_{\rm f} = N_{\rm b} = L/2$
(see Fig.~\ref{Sk}).

On the contrary, in the FM regime, we see no sign of oscillations and
observe only a slow FM relaxation. For the special case where $J_{\rm
z} = 0$ ($U_{\rm bf} = U_{\rm bb}$) the system is mapped on a pure
$xy$ pseudo-spin Hamiltonian. The AF and FM correlation functions are
expected to decay with the same power law \cite{Giam} and we observe
the combined effect of both types of correlation (see Fig.~\ref{Sk}).

In the SFb phase we have already seen that $xy$ correlations are
present (Fig.~\ref{green}). It is more difficult to observe the PSDW
correlations along the $z$ axis.  To obtain the SFb phase, one needs
$U_{\rm bf} > U_{\rm bb}$ for the Mott phase to be destroyed. This
means that the filled layer is in the FM regime. However, it is
difficult to distinguish the FM signal due to the filled layer from
the signal due to the superfluid boson density-density correlations
which are expected to show some kind of short range relaxation. On the
contrary, an AF signal would be due to the filled layer only, but
should be small because we are in the region where the dominant
effects are FM.  We observe such a weak AF signal with a decreasing
amplitude when $U_{\rm bf}$ increases, as expected (Fig.~\ref{Sk},
inset).

Finally the interpretation of the SFb as composed of superfluid bosons
on a filled layer in a pseudo-spin phase is partly supported by our
data.  We observe, as expected, that the leading correlations are the
bosonic Green functions followed by the composite Green function.  We
also found that the measure of the pseudo-spins correlations along $z$
are consistent with the PSDW phase.  However, the decay of the
fermionic Green function is algebraic which is consistent with the
non-dominant PSDW correlations. The present algorithm gives access to
one- and two-body correlation functions and topological information
via the winding numbers. It does not give information about other
$N-$point correlation functions or what are the dominant
correlations.

In conclusion, we have studied the effect of adding fermions on a
boson Mott insulator and varying the interaction strength between
fermions and bosons. We observed that, when the Mott insulator is
destroyed, the system enters a superfluid phase where the movement of
bosons and fermions is correlated.  The similarities between this
phase and a previously observed pseudo-spin phase led us to describe
it, approximately, as a phase displaying simultaneously traditional
superfluidity and PSDW.

Even in non symmetric cases, where the number of bosons and fermions
is not equal, it is possible to observe some of the characteristics of
the original pseudo-spin phase as long as interactions are large.
These results lead us to conclude that anticorrelated motion of bosons
and fermions is a robust property and appears generally in strongly
interacting mixtures of bosons and fermions.

The authors thank U. Schollw\"ock, B. Svistunov, and M. Troyer for
useful discussions.  We acknowledge financial support from a grant
from the CNRS (France) PICS 18796 and from the Swiss National Science
Foundation. Part of the calculations were carried out on the
Hreidar-cluster at ETH Zurich.


\begin{thebibliography}{10}
\bibitem{LL} These two regions are essentially the same as the two 
Luttinger's liquid regions (LLa and LLb) found in \cite{Pollet1}
\end{thebibliography}
\end{document}